# Source-Filter Decomposition of Harmonic Sounds

Ilia Bisnovatyi and Michael O'Donnell
Department of Computer Science, University of Chicago
1100 E. 58th street, Chicago IL 60637, USA
ilia@cs.uchicago.edu, odonnell@cs.uchicago.edu

**Abstract**

This paper describes a method for decomposing steady-state instrument data into excitation and formant filter components. The input data, taken from several series of recordings of acoustical instruments is analyzed in the frequency domain, and for each series a model is built, which most accurately represents the data as a source-filter system. The source part is taken to be a harmonic excitation system with frequency-invariant magnitudes, and the filter part is considered to be responsible for all spectral inhomogenieties. This method has been applied to the SHARC database of steady state instrument data to create source-filter models for a large number of acoustical instruments. Subsequent use of such models can have a wide variety of applications, including improvements to wavetable and physical modeling synthesis, high quality pitch shifting, and creation of "hybrid" instrument timbres.

## 1  Introduction

Digital simulation of sounds produced by acoustical instruments has been an important goal of digital synthesis algorithms throughout the history of their development. The ability to faithfully reproduce "natural instruments" has often been viewed as both an important achievement in itself and a good starting point for developing new synthetic timbres. Such importance arises from the fact that the aesthetic perception of timbre is shaped to a great extent by listeners' exposure to the sounds of traditional acoustical instruments and human voice. The combined memory of sounds that can be produced by an instrument throughout its playing range and via a set of applicable playing techniques amounts to a perceptual concept of that instrument and aids in the identification and aesthetic placement of that instrument in a greater musical context.

One could argue that the perceptual properties of a musical note can be roughly subdivided into two categories, namely, spectral and dynamic or temporal characteristics. The former would include pitch and harmonicity, while the latter would include note envelope (attack, steady state, and decay durations), overall timbral development (e.g. late rise of higher partials in brass instruments), etc. When an instrument is played, these properties can be varied in a number of ways – some may be altered through different playing techniques, while others change with pitch. In order to produce a playable and flexible musical instrument by means of digital synthesis, one needs to have a good understanding of mappings from the control parameters of a particular algorithm to the perceptual qualities of the resulting timbre[1]. It is also important to identify and reproduce the properties that will remain invariant throughout a set of playing techniques and/or throughout the entire pitch range of the target instrument.

Such considerations form a large part of the motivation behind the research in physical modeling techniques. One expects that by developing a synthesis method modeled after physical processes that take place in a real instrument, one should be able to produce control structures that will correspond to "natural" control parameters operated upon by a player of the original instrument, as well as better deal with traditionally difficult aspects of synthesis, such as transitions between notes and extended playing techniques. While these attempts have been partly successful, controlling physical models is a fairly difficult task. A high level of model complexity, necessary for algorithm flexibility and faithful reproduction of physical processes, leads to a large set of parameterized elements in a model. As a result, one is left with the arduous task of finding a suitable mapping from a small space of controls accessible to the user to a large space of model parameters. Such mapping may end up being highly non-linear and time-variant.

---

[1] There are, of course, methods of synthesis that succeed, to a degree, while providing only a small subset of such mappings. However, these methods usually have either limited or extremely non-intuitive control parameters.

## 2 Intermediate Representations and Source-Filter Model

A possible improvement on pure physical modeling can be achieved through the use of intermediate representations, which one may call *physically informed* methods. Such methods will have some degree of physical information encoded in them, but will not attempt to achieve precision[2] in the model. Where it appears advantageous, for reasons of computational efficiency or ease of control, these methods will rely on "physically uninformed" numerical and/or statistical techniques, much like wavetable synthesis relies on large amounts of stored numerical data, without having any "physical understanding" of it. The hope is that by carefully designing such an algorithm, one would be able to achieve a reduction of both data and computational complexity, while providing a simpler control structure.

This paper describes an algorithm which facilitates the design of such intermediate representations based on a *source-filter* model. This model has been employed extensively in speech research [1], and has been integrated into both physical modeling methods [3] and sampling, or wavetable synthesis [4]. The basic motivation behind the use of this model is that most natural instruments, as well as human voice can be thought of as consisting of two fundamental parts - a sound producing excitation system and a filter, which determines the overall spectral character of the instrument.[3] Another important factor is that a source-filter model facilitates independent control of dynamic and spectral properties of synthesized sounds.

In order to successfully construct source-filter models for specific instruments, one needs to obtain numerical data for the source and filter components. A physical modeling approach would involve obtaining the physical measurements of the materials that the instruments is made of, their precise geometry, modes of vibration, etc. While there are some formidable examples of in-depth study of the physical properties of specific instruments (notably, [7]), presently such implementation would be practically and computationally out of reach. The approach proposed in this paper will forgo physical modeling of the two parts in favor of obtaining their numerical models by means of statistical analysis of the sounds produced by the target instrument. This approach has the advantage of drastically reduced complexity and can be easily adapted to a large variety of different instruments. All that is required is that the instrument be nearly harmonic and that a recording of every note in its range be available. The algorithm analyzes the series of recordings and attempts to find a source-filter decomposition that most closely represents the original data.

## 3 Input Data

The algorithm described here is performed in the frequency domain; for each note in the playing range, the spectrum of the steady-state portion of the sound is used as input data. A practical implementation has been tested out on SHARC, a timbral database covering a large number of acoustical instruments, which is free and is readily available on the web [2]. SHARC contains steady state data for a large number of acoustic instruments, which makes it very well suited for our purposes. For each instrument, a chromatic series of notes has been analyzed (every note had been individually played and digitally recorded; a description of the original recordings can be currently found in [6]). Several periods of steady-state sound had been selected from each note in a series and spectrally analyzed. SHARC assumes that all input sounds are harmonic, and since for every note the fundamental is known, the steady-state data can be represented as a set of values for magnitudes and phases of partials. The total number of detected partials varies from note to note, and the range and total number of notes varies from instrument to instrument. For the purposes of uniformity, we choose to consider the same number of partials for every sample; higher notes will tend to have fewer partials due to the Nyquist limit, therefore one can either disregard some of the partials for the lower notes, or truncate the input series. It is possible to fine tune the algorithm by varying the constraints on the minimum number of samples and/or partials that is required before truncation is allowed (see *Appendix B* for examples).

While both phase and magnitude data are required for full filter reconstruction, this paper will focus on processing the magnitude data. Phase information can be obtained independently in a nearly identical

---

[2] *Precision* here is understood to mean proximity of the model to the physical instrument, rather than numerical accuracy.

[3] As an example, consider a string instrument, such as cello, as consisting of a string-bow-neck system, that acts as an excitation and an instrument body, which is a filter.

fashion (section 7 outlines the specifics of dealing with phase).

## 4  Representation

Let $S$ be the total number of chromatic samples in a series, and $K$ – the smallest number of available partials for any given sample (for the purposes of uniformity, we choose to consider the same number of partials for every sample; higher notes will tend to have fewer partials due to the Nyquist limit, therefore one can either disregard some of the partials for the lower notes, or truncate the input series). Let $D_{i,j}$ be the amplitude of $j$-th partial of $i$-th note – these are the data points. Now consider an equally spaced grid in the log frequency space, whose bins are centered on the fundamentals of equally tempered chromatic tones. This grid will define the resolution for the formant filter coefficients $R_n$, i.e. for each bin the magnitude of filter's frequency response in that bin will have to be determined. This resolution is reasonable, because the formant curve is expected to be fairly smooth and because for most traditional applications one will rarely need to synthesize notes less than a semitone apart (however, if required, an interpolated curve can be used). Note also that this resolution is only determined by the spacing of the original samples, and adapting to a more finely sampled input would be trivial. The target excitation system will consist of $K$ partials with amplitudes $P_j$, which remain constant for every sample. Additionally, to account for the differences in the musical performance of individual notes, an overall multiplicative scaling coefficient $A_i$ for each sample is introduced.

The data points and the variables are related by a set of equations

$$D_{i,j} = A_i P_j R_n \qquad (1)$$

for $i=1..S$ and $j=1..K$. Index $n$ is the number of the bin into which the frequency of the $j$-th partial of $i$-th sample falls, starting with 1 for the fundamental of the lowest note, i.e. $n = \lfloor 12 \log_2 j + 1/2 \rfloor + i$.

All of the values in (1) are positive, and thus, to facilitate the solution, the products in (1) can be easily converted into sums by switching to a logarithmic magnitude scale:

$$d_{i,j} = a_i + p_j + r_n \qquad (2)$$

where $d_{i,j} = \ln(D_{i,j})$, $a_i = \ln(A_i)$, $p_j = \ln(P_j)$, and $r_n = \ln(R_n)$. This is a system of $S*K$ linear algebraic equations; the data matrix $\{d_{i,j}\}$ can be collapsed into a vector $\bar{d}$, and all the variables – into a vector $\bar{v}$, thus transforming (2) into a linear system

$$\bar{v}\,\mathrm{M} = \bar{d}, \qquad v_k = \begin{cases} p_k, & k \leq K \\ a_{k-K}, & K < k \leq S+K \\ r_{k-S-K}, & k > S+K \end{cases} \qquad (3)$$

where M is the corresponding matrix of zeroes and ones. *Figure 1* shows a graphical representation of M for $S=12$, and $K=16$, with ones marked in black :

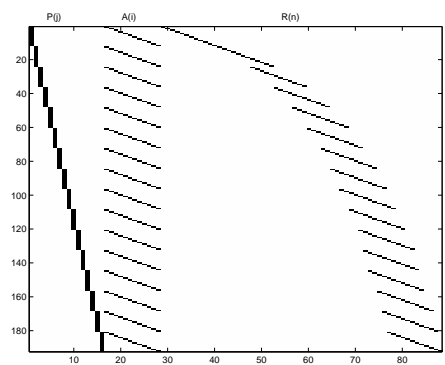

Fig.1

## 5  Approaches to the solution

The system (3) is generally underdetermined, since the rank of M is always less than $S+K+N$. One extra degree of freedom can be easily eliminated – an overall scaling factor that could be applied to the excitation at the expense of scaling coefficients $A_i$. However, even after normalizing the excitation (setting $p_0=0$ and eliminating the first column from M), the system will remain underdetermined (for all practically interesting cases this can be verified empirically by computing the rank of M).

There are many ways in which constraints could be added to (3) in order to choose the solution. For example, assumptions could be made about the smoothness of the filter or about the range into which the scaling coefficients $\{A_i\}$ fall. For the general case, after some experimentation, a robust iterative method was chosen. The iterations alternate between solving for $\{p_j\}$ given $\{r_n\}$ and solving for $\{r_n\}$ given [normalized] $\{p_j\}$. No special assumption is made about the values of $a_i$ – they are readjusted after each iteration. A weighted least-squares convergence metric is used as a test for the termination of the iterative process. For every instrument from SHARC this algorithm converges within 20 iterations, allowing for deviations of <0.1%.

## 6 Results

The figures in *Appendix A* illustrate some of the strengths and shortcomings of the proposed method. The pairs of figures 2,3 and 4,5 show the excitation and filter solutions for plucked and bowed cello respectively ($S$=28, $K$=32, $F_0$=65.406Hz). As one would expect, the excitations are somewhat different, while the filter curves exhibit similar resonant properties, although they are not identical. Figures 6 and 7 show the excitation and filter solutions for bass clarinet ($S$=25, $K$=32, $F_0$=69.296Hz). The suppression of even partials is clearly evident in the excitation, which conforms to the physical process of harmonic generation in clarinets [5].

## 7 Towards full reconstruction

As was mentioned previously, one needs to process the phase information in order to reconstruct the filter entirely. Generally speaking, phase data can be subjected to the same treatment as the magnitude data. However, one needs to be aware of phase rollover, since all phase data is *mod* $2\pi$, which leads to extra degrees of freedom and may require additional constraints on the choice of the solution. Another potential difference between phase and magnitude is that the phases of partials may be subject to specific constraints. The most obvious example would be the assumption of phase-locking in the excitation, which corresponds to our understanding of the natural processes occurring in human voice and some instruments. This would mean that the excitation system will have only one phase variable for each sample -- an overall phase shift, while the filter will be responsible for all phase deviations of the partials. On the other hand, if the actual partials are not precisely harmonic, their deviations may, depending on the method of frequency analysis, show up as phase shifts.

Once the phase information has been obtained, one can apply the complete filter model to the entire sample data (via deconvolution) in order to obtain the time-trajectories of the partials during attacks and other non-steady-state portions of the sound. This data can be subsequently used for artifact-free time-stretching and pitch shifting transforms, as well as for creation of "hybrid" instruments.

## 8 Summary and discussion

The algorithm described here provides a fast and simple tool for obtaining excitation and filter components from steady-state data. The problem is reduced to a system of linear equations, which is generally under-constrained, and an iterative solution method has been proposed, which, we believe, selects qualitatively appropriate solutions. The final representation of the original magnitude data is precise; there is no data loss. An automated interface for the SHARC database has been built, providing excitation and filter patterns for a large number of acoustical instruments.

There are several directions for further improvement. As was mentioned previously, applying carefully selected constraints to the variables can lead to a more appropriate choice of solution. With a slight modification, the algorithm could collect more information in cases when different excitation patterns are processed by the same filter (such as recordings of the same instrument played via different techniques) or the same excitation applied to different filters (e.g. a voice singing different vowels). Similarly, more information can be obtained by analyzing the same series played a number of times, since repetition will tend to average out the effects of uneven performances.

Currently, the authors' work in this area is focused on determining the optimal ways for integrating phase information and on developing the resynthesis/transformation framework based on data obtained from source-filter analysis.

# *Appendix A*

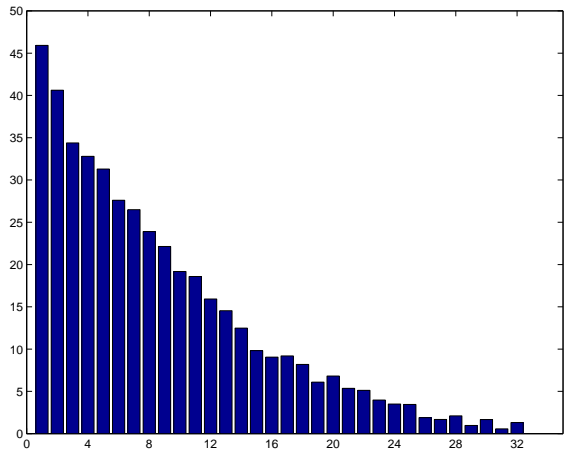

Fig.2   Cello pizzicato (excitation).

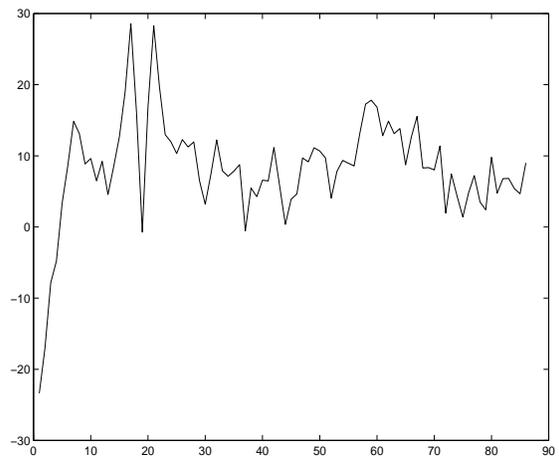

Fig.3   Cello pizzicato (filter).

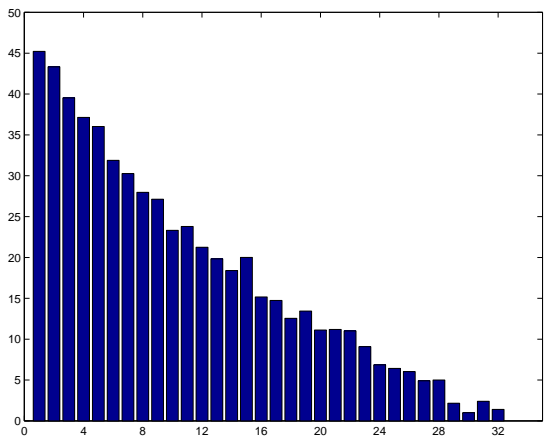

Fig.4   Cello martele (excitation).

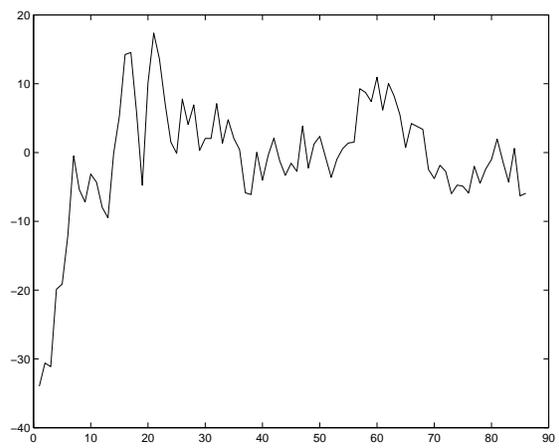

Fig.5   Cello martele (filter).

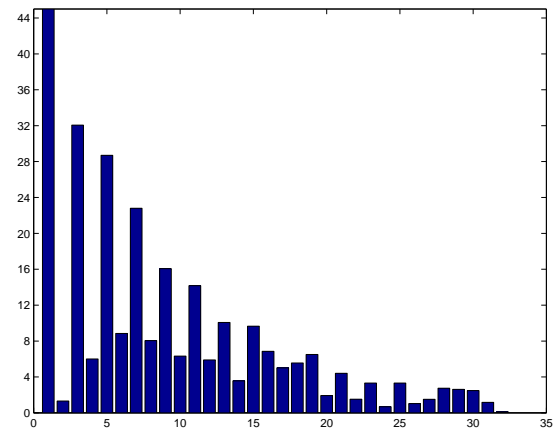

Fig. 6   Bass clarinet (excitation).

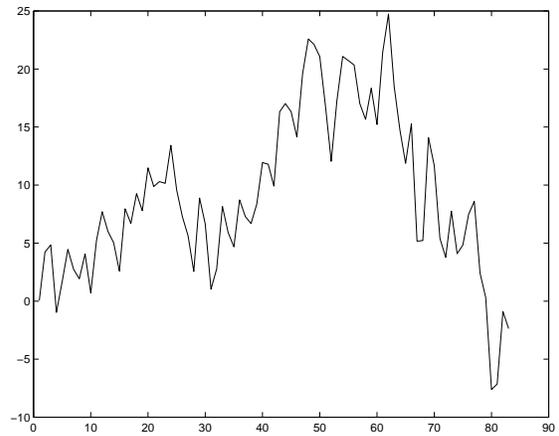

Fig.7   Bass clarinet (filter).

## Appendix B
## SHARC instruments and analysis parameters

| Instrument | $F_0$ (Hz) | $S_1$ | $K_1$ | $S_2$ | $K_2$ |
|---|---|---|---|---|---|
| Bach trumpet | 246.942 | 17 | 16 | 29 | 8 |
| Bb clarinet | 146.832 | 27 | 16 | 38 | 8 |
| Bass | 32.703 | 39 | 36 | | |
| Bass martele | 32.703 | 38 | 38 | | |
| Bass muted | 32.703 | 40 | 33 | | |
| Bass pizzicato | 32.703 | 42 | 27 | | |
| C trumpet | 184.997 | 23 | 16 | 34 | 8 |
| C trumpet muted | 184.997 | 23 | 16 | 32 | 9 |
| Eb clarinet | 195.998 | 22 | 16 | 33 | 8 |
| English horn | 164.814 | 25 | 16 | 31 | 11 |
| French horn | 73.416 | 38 | 16 | | |
| French horn muted | 73.416 | 38 | 17 | | |
| Bassoon | 58.270 | 33 | 28 | | |
| Alto trombone | 349.228 | 14 | 14 | | |
| Alto flute vibrato | 195.998 | 22 | 16 | 31 | 9 |
| Bass clarinet | 69.296 | 26 | 36 | | |
| Bass trombone | 43.654 | 26 | 57 | | |
| Bass flute vibrato | 130.813 | 27 | 18 | | |
| Cello martele | 65.406 | 40 | 17 | 45 | 12 |
| Cello muted vibrato | 65.406 | 35 | 22 | | |
| Cello pizzicato | 65.406 | 40 | 17 | | |
| Cello vibrato | 65.406 | 41 | 16 | 45 | 12 |
| Contrabass clarinet | 46.249 | 24 | 60 | | |
| Contrabassoon | 29.135 | 33 | 57 | | |
| Oboe | 233.082 | 18 | 16 | 30 | 8 |
| Flute vibrato | 261.626 | 16 | 16 | 29 | 8 |
| Piccolo | 587.330 | 16 | 7 | 26 | 4 |
| Trombone | 82.407 | 37 | 16 | | |
| Trombone muted | 82.407 | 34 | 19 | | |
| Tuba | 65.406 | 32 | 25 | | |
| Viola martele | 130.813 | 28 | 16 | 37 | 9 |
| Viola muted vibrato | 130.813 | 28 | 16 | 40 | 8 |
| Viola pizzicato | 130.813 | 29 | 16 | 35 | 11 |
| Viola vibrato | 130.813 | 29 | 16 | 40 | 8 |
| Violin martele | 195.998 | 22 | 16 | 34 | 8 |
| Violin muted vibrato | 195.998 | 21 | 17 | 34 | 8 |
| Violin pizzicato | 195.998 | 22 | 16 | 34 | 8 |
| Violin vibrato | 195.998 | 21 | 17 | 33 | 8 |

--------------------
The above table lists the SHARC instruments for which the analyses were performed with their corresponding parameters. The second column shows the lowest available fundamental. $S_1$ and $K_1$ are the numbers of available samples and partials with truncation of the series permitted after 16 samples when the number of partials falls below 16. $S_2$ and $K_2$ show those numbers when truncation is permitted after 24 samples with the number of partials falling below 8. Where $S_2$ and $K_2$ are omitted, they are identical to the corresponding $S_1$ and $K_1$.